\documentclass[useAMS,usenatbib,usegraphicx]{mn2e}

\title[Ratios of Star Cluster Radii]{Ratios of star cluster core and half-mass radii: a cautionary note 
on intermediate-mass black holes in star clusters}
\author[J. R. Hurley]{Jarrod R. Hurley$^{1,2}$\thanks{
E-mail: jhurley@swin.edu.au (JRH)} \\
$^{1}$Centre for Astrophysics and Supercomputing, Swinburne University of Technology, P.O. Box 218, VIC 3122, Australia \\
$^{2}$Department of Astrophysics, American Museum of Natural History, 
            Central Park West at 79th Street, New York, NY 10024, USA}

\begin{document}

\date{Accepted 2007 Month xx. Received 2007 Month xx; in original form 2007 March 26} 

\pagerange{\pageref{firstpage}--\pageref{lastpage}} \pubyear{2005}

\maketitle

\label{firstpage}

\begin{abstract}
There is currently much interest in the possible presence of intermediate-mass 
black holes in the cores of globular clusters. 
Based on theoretical arguments and simulation results it has previously been 
suggested that a large core radius -- or particularly a large ratio of the core radius 
to half-mass radius -- is a promising indicator for finding such a 
black hole in a star cluster. 
In this study $N$-body models of $100\,000$ stars with and without primordial 
binaries are used to investigate the long-term structural evolution of star clusters. 
Importantly, the simulation data is analysed using the same processes by 
which structural parameters are extracted from observed star clusters. 
This gives a ratio of the core and half-mass (or half-light) radii that is directly comparable 
to the Galactic globular cluster sample. 
As a result, it is shown that the ratios observed for the bulk of this sample can be explained 
without the need for an intermediate-mass black hole. 
Furthermore, it is possible that clusters with large core to half-light radius ratios
harbour a black-hole binary (comprised of stellar mass black holes) 
rather than a single massive black hole. 
This work does not rule out the existence of intermediate-mass black holes in 
the cores of at least some star clusters. 
\end{abstract}


\section{Introduction}
\label{s:intro}

The situation regarding the growing body of evidence that some globular clusters (GCs) 
may be harbouring intermediate-mass black holes (IMBHs) has been summarized 
recently by Baumgardt, Makino \& Hut (2005). 
This evidence includes taking the relationship found between the masses of supermassive 
black holes (BHs) and the bulge masses of the host galaxies (Magorrian et al. 1998) and 
extrapolating to globular cluster masses (Kormendy \& Richstone 1995). 
For a typical globular cluster, such as M15 (van der Marel 2001), this gives a BH mass 
of $\sim 10^3 \, M_\odot$. 
Sitting conveniently between the supermassive and stellar-mass BH regimes 
-- where the latter includes BHs of $\sim 50 \, M_\odot$ or less -- 
the IMBH tag arises naturally. 
The existence of such BHs is backed up by the $N$-body simulations of 
Portegies Zwart et al. (2004) showing that possible progenitors 
(main-sequence stars of $\sim 10^3 \, M_\odot$) can be created through runaway 
mergers of massive stars in young clusters. 
Detection is possible through the measurement of central velocity dispersions 
in globular clusters but  this is a challenging process 
(Baumgardt, Makino \& Hut 2005; Trenti 2006). 
To date this has led to suggestions of an IMBH in the core of M15 (Gerssen et al. 2002) 
and in the core of G1 (Gebhardt, Rich \& Ho 2002). 
However, Baumgardt et al. (2003a, 2003b) subsequently used $N$-body simulations 
to show that the inferred non-luminous central mass could instead be a central 
concentration of stellar-mass BHs, white dwarfs and neutron stars 
(but see also Gebhardt, Rich \& Ho 2005). 

Notwithstanding the lack of direct confirmation that IMBHs do reside in the 
cores of GCs, study into the ramifications of such a scenario has progressed.  
Importantly, Baumgardt, Makino \& Hut (2005) have shown that a GC with an IMBH 
in the core will be observed to have a relatively flat central surface brightness profile 
and consequently a larger measured core-radius compared to a GC without an IMBH. 
This result has been followed up by Trenti (2006) who suggests that the ratio of the 
core radius, $r_{\rm c}$, to the half-mass radius, $r_{\rm h}$, of a dynamically-evolved 
cluster can be used to infer the presence of an IMBH. 
Trenti (2006) combines results from a variety of $N$-body simulations 
(Heggie, Trenti \& Hut 2006; Trenti, Heggie \& Hut 2007; Trenti et al. 2007).  
These show that $r_{\rm c} / r_{\rm h} \sim 0.02$ for clusters composed initially of 
single stars only, $r_{\rm c} / r_{\rm h} \sim 0.05$ for clusters with primordial binaries, 
and $r_{\rm c} / r_{\rm h} \sim 0.3$ for clusters with an IMBH. 
These values are taken when the model clusters are relaxed systems and 
the core-collapse phase has ended. 
In comparison, observations of Galactic GCs show a distribution of $r_{\rm c} / r_{\rm h}$ 
extending from $0.1 - 1.0$ with a peak at about 0.5 (Fregeau et al. 2003). 
From a theoretical viewpoint Heggie et al. (2006) examine how the $r_{\rm c} / r_{\rm h}$ 
ratio varies with the BH mass. 
This also suggests that a star cluster observed to have a large core radius 
presents the most promising target for finding an IMBH, in the sense that 
large mass implies large core radius. 
As with the above results this argument is only valid in the post-collapse regime. 

A recurring issue with $N$-body simulations of star cluster evolution is that the 
models are generally idealized in some way (or ways) that prohibits direct 
comparison to real clusters. 
The simulations of Heggie, Trenti \& Hut (2006), Trenti, Heggie \& Hut (2007) 
and Trenti et al. (2007) were restricted to initial particle numbers of $N_0 = 20\,000$ or less 
and assumed equal-mass stars. 
As pointed out by Trenti (2006) these results can be scaled to GC particle 
numbers ($N_0 \sim 10^5 - 10^6$) but only by also neglecting stellar evolution. 
Simulations performed by 
Baumgardt \& Makino (2003) and Baumgardt, Makino \& Ebisuzaki (2004) 
included particle numbers up to $131\,072$ stars, a mass spectrum and stellar evolution. 
However, primordial binaries were not included. 
Another key factor is that one must be sure to compare like-with-like when using 
model and real data. 
Specifically this relates to use of the core radius, half-mass (or half-light) radius, 
and the half-mass relaxation timescale, $t_{\rm rh}$. 

Considering the growing interest in IMBHs it is only natural that attempts are being 
made to isolate key observational tests for their existence. 
Unfortunately, in this paper, it is shown that $r_{\rm c} / r_{\rm h}$ cannot readily be used 
as such a test. 
This is based on a series of $N$-body models of $100\,000$ stars with and 
without primordial binaries. 
The models include a full mass spectrum, stellar and binary evolution, and 
account for the tidal field of the Galaxy. 
The models do not include IMBHs. 
Model data is analysed using a pipeline analogous to that used to reduce 
real cluster data. 

Section~\ref{s:models} gives a description of the models used in this 
work including the initial setup of the models and an overview of the evolution. 
A detailed look at the internal structure of the model clusters is then given in 
Section~\ref{s:results} along with a description of the attempt to analyse model data as 
real data. 
This is followed by a discussion in relation to previous work and observations 
of Galactic GCs, and finally a summary of the main results.

\section{Models}
\label{s:models}

The focus of this work is a set of realistic $N$-body simulations that each 
starts with $N = 100\,000$ objects 
-- an object being either a star or a binary. 
Specifically, the starting models contain: 
$100\,000$ single stars and no primordial binaries (labelled the K100-00 simulation); 
$95\,000$ single stars and $5\,000$ binaries (K100-05); and, 
$90\,000$ single stars and $10\,000$ binaries (K100-10). 
Masses for the stars are chosen from the initial mass function (IMF) 
of Kroupa, Tout \& Gilmore (1993) between the limits of $0.1 - 50 \, M_\odot$. 
Metallicity is set at $Z = 0.001$ for the stars. 
The initial positions and velocities are assigned according to a Plummer 
density profile (Plummer 1911; Aarseth, H\'{e}non \& Wielen 1974) in virial equilibrium. 
A scale length of $8.5\,$pc is set for each simulation 
-- this is to comply with the tidal radius set by the external tidal field (see below). 
In actual fact the results from two simulations starting with $100\,000$ single stars 
will be utilised. 
These simulations are identical in all respects except for the random number seed used to generate 
the starting masses, positions and velocities. 
These will be known as K100-00a and K100-00b. 
See Table~\ref{t:table1} for a list of the simulations used in this work. 

The model clusters are evolved using the  {\tt NBODY4} code (Aarseth 1999, 2003). 
This includes algorithms for stellar and binary evolution as described in Hurley et al. (2001). 
Simulations are performed using 
32-chip GRAPE-6 boards (Makino 2002) located at the 
American Museum of Natural History. 
Each simulation took approximately six months to complete on a dedicated 
GRAPE-6 board. 

To account for the tidal field of the Galaxy each cluster is placed on a circular orbit 
at a distance of $8.5\,$kpc from the Galactic centre with an orbital speed of 
$220\, {\rm km} {\rm s}^{-1}$. 
This is commonly referred to as a Standard Galactic tide (see Giersz \& Heggie 1997 
for a full description). 
For the model clusters in this work, which each have a starting mass of $M \sim 50\,000 \, M_\odot$, 
this gives an initial tidal radius of about $50\,$pc. 
With the length-scale given above the clusters are close to filling their tidal radii 
at birth, noting that the position of the outermost star will vary from model to model 
as positions are drawn at random from a distribution. 

Each cluster was evolved to a minimum age of $16\,$Gyr. 
This ensured that the core-collapse phase of evolution was completed and that models of 
comparable age to GCs were available for analysis. 
In fact, for model K100-00b it is not necessarily true that core-collapse was reached. 
For reasons that will become evident in Section~\ref{s:results} this model did not 
show a deep minimum in core-radius prior to its termination at $16\,$Gyr 
whereas the other three models did show such a minimum between $15-16\,$Gyr. 
For interest sake the K100-05 simulation was allowed to proceed to $20\,$Gyr. 
After $16\,$Gyr of evolution the model clusters had been reduced to $N \sim 22\,000$ 
and, in terms of mass, approximately 80\% of the cluster had been lost over that period. 
The tidal radius at $16\,$Gyr was about $30\,$pc. 

\begin{figure}
\includegraphics[width=84mm]{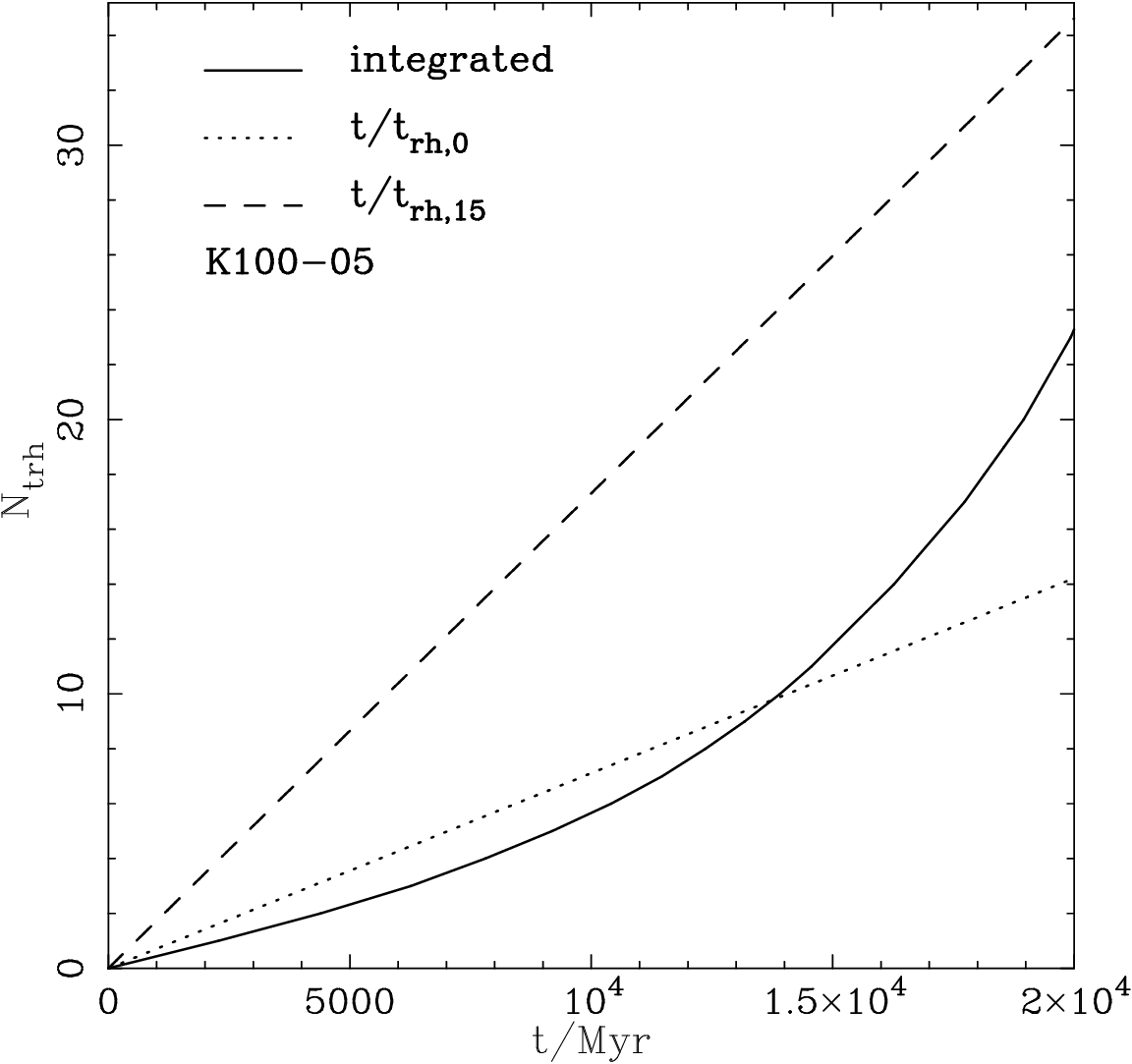}
\caption{
A comparison of methods used for calculating the dynamical age of a star cluster in 
terms of the number of half-mass relaxation times elapsed as a function of time. 
Shown is the age of the cluster simply divided by either the initial half-mass relaxation timescale 
($t_{\rm h,0} = 1\,400\,$Myr: dotted line) or the half-mass relaxation timescale at an 
age of $15\,$Gyr ($t_{\rm h,0} = 580\,$Myr: dashed line). 
These are compared to the more detailed method of accumulating, or integrating, the 
number of half-mass relaxation times elapsed as the model evolves (solid line). 
Data is from the simulation starting with $95\,000$ single stars and $5\,000$ binaries. 
\label{f:fig1}}
\end{figure}

The evolution of the K100-05 model is shown in Figure~\ref{f:fig1} in terms of the 
number of half-mass relaxation times that have elapsed. 
This is done using both the initial half-mass relaxation timescale 
($t_{\rm rh,0} = 1\,400\,$Myr) and the timescale after $15\,$Gyr ($t_{\rm rh,15} = 580\,$Myr). 
The difference between the two is significant  and shows that one must be very 
careful using the observationally determined $t_{\rm rh}$ of a cluster to infer the 
dynamical age (this point will be returned to later). 
The relaxation time is calculated according to the standard expression developed 
by Spitzer (1987: see eq.~1 of Baumgardt, Makino \& Hut 2005). 
In reality $t_{\rm rh}$ is an evolving quantity, generally decreasing with age, and this 
must be accounted for when calculating the true dynamical age 
(see the solid-line in Figure~\ref{f:fig1}). 
The evolution of $t_{\rm rh}$ for the K100-00 and K100-10 simulations differs from 
the K100-05 simulation by no more than a few per cent across the evolution.

\section{Results}
\label{s:results}

\begin{figure*}
\includegraphics[width=168mm]{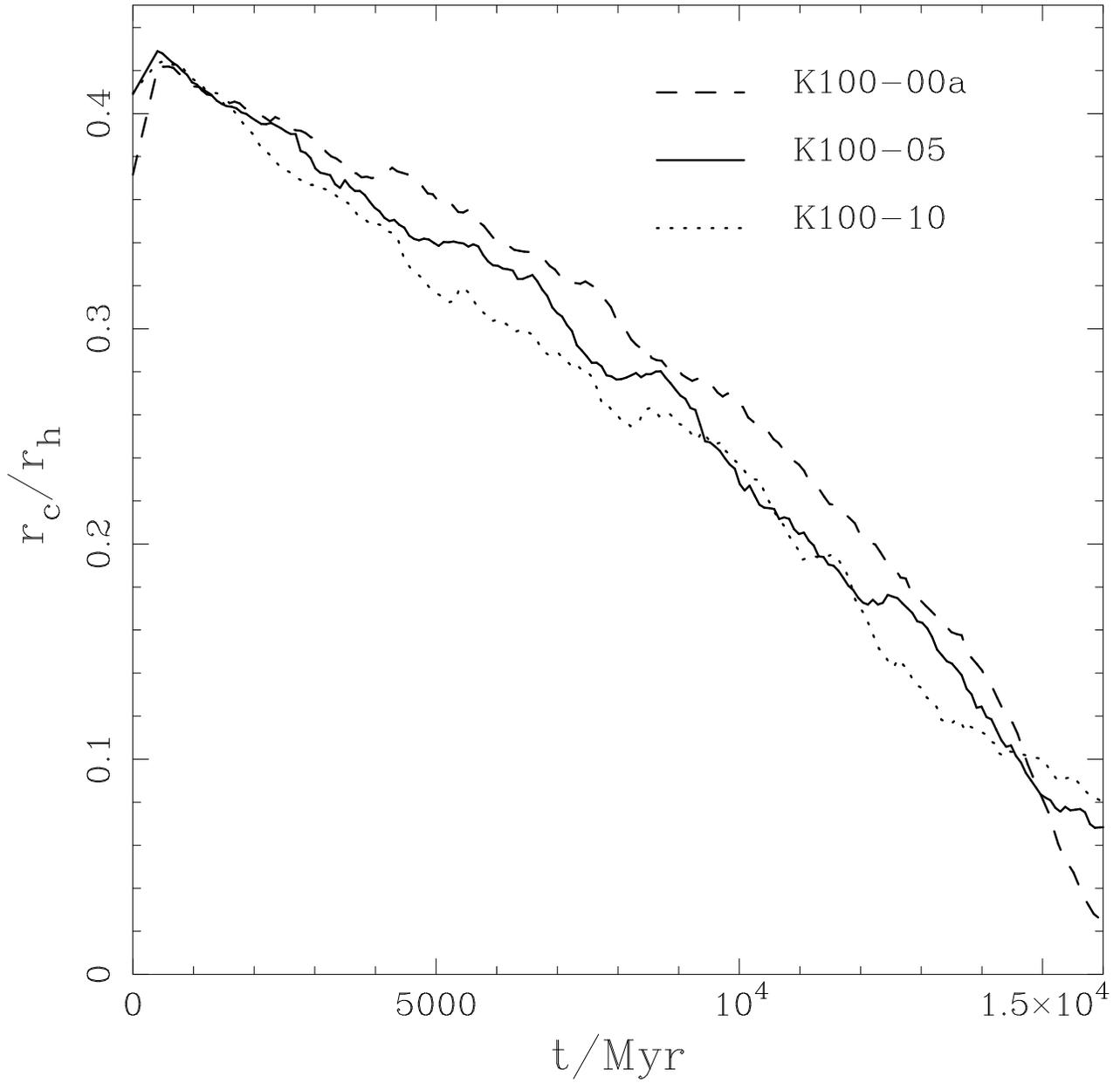}
\caption{
Evolution of the ratio of the core-radius, $r_{\rm c}$, to half-mass radius, $r_{\rm h}$, 
for models starting with 0, 5 and 10\% binaries (see Table~\ref{t:table1} for a description). 
The radii are calculated using standard $N$-body methods and three-dimensional 
data (see text for details). 
\label{f:fig2}}
\end{figure*}

The evolution of $r_{\rm c} / r_{\rm h}$ for the models starting with 0\%, 5\% and 10\% 
binary frequency is shown in Figure~\ref{f:fig2}. 
Here $r_{\rm c}$ is the density-weighted core-radius (Casertano \& Hut 1985) 
commonly used in $N$-body simulations and $r_{\rm h}$ is the half-mass radius. 
These are not directly comparable to observed quantities. 

Initially $r_{\rm c} / r_{\rm h}$ increases for all models. 
This is because the early phase corresponding to rapid mass-loss 
from massive stars leads to an overall expansion and the effect is greater at smaller radii. 
Subsequent evolution has $r_{\rm c} / r_{\rm h}$ generally decreasing as it is dominated 
by the contracting core -- $r_{\rm h}$ continues to expand until about $4\,$Gyr.  
After that time the half-mass radius begins to feel the effect of the decreasing tidal radius 
and gradually decreases from that point on. 

The evolution of $r_{\rm c} / r_{\rm h}$ is similar for all models at all times. 
At the $16\,$Gyr end-point there is some distinction between the models 
with and without primordial binaries: $r_{\rm c} / r_{\rm h} \sim 0.07$ in the former 
and $\sim 0.02$ in the latter. 
However, the data in Figure~\ref{f:fig2} have been smoothed considerably 
using a moving $500\,$Myr window and $100\,$Myr increments. 
In reality the noise in the data would preclude drawing any inference regarding the 
primordial binary content of a cluster based on its $r_{\rm c} / r_{\rm h}$ measurement. 
This is not to say that a systematic difference in core radius would not develop if 
the models were allowed to evolve well in to the post-core-collapse regime. 

In terms of comparing to real data the results of Figure~\ref{f:fig2} are not particularly 
useful. 
What is needed is a procedure that analyses the model data in the same way as is 
done for observations of clusters. 
In this way a meaningful $r_{\rm c} / r_{\rm h}$ ratio can be extracted. 
The $N$-body stellar evolution algorithm (Hurley, Pols \& Tout 2000) provides the mass, 
luminosity and effective temperature of each model star. 
Using the model atmosphere data of Kurucz (1992), supplemented by Bergeron, 
Wesemael \& Beauchamp (1995) for white dwarfs, these are then converted to 
broadband UVBRI colours. 
It is then relatively simple to calculate the half-light radius, $r_{\rm h,l}$ as the 
radius which encompasses the inner half of the total light of the cluster. 
This is a projected radius calculated using a 2-dimensional projection of 
the 3-dimensional positions of the model stars. 
Finding the observational core-radius, which will be labelled $r_{\rm c,l}$, requires 
analysis of the cluster surface brightness profile (SBP). 
For this it is possible to use the software described by Mackey \& Gilmore (2003) in their work 
on the star clusters of the Large Magellanic Clouds. 
Each $N$-body snapshot is taken in turn and used to construct a two-dimensional projected SBP. 
Stars more than two magnitudes brighter than the main-sequence turn-off and low-mass 
stars with $M_{\rm V} > 10$ are excluded -- this mimics the observational process of 
avoiding bright stars, which may saturate, and faint stars which may be incomplete in 
number. 
Note that the projection is taken along the Y-axis and a choice is made to 
focus on the V magnitude. 
Neither of these choices affects the results to any significant degree. 
Next a three-parameter Elson, Fall \& Freeman (EFF: 1987) model is fitted to the 
cluster SBP to determine $r_{\rm c,l}$ (Mackey \& Gilmore 2003). 
A similar approach was taken by Heggie et al. (2006) although the fit was made to 
the three-dimensional density profile and a 
fourth parameter was added in order to fit the central cusp for models with a central BH. 

\begin{figure*}
\includegraphics[width=168mm]{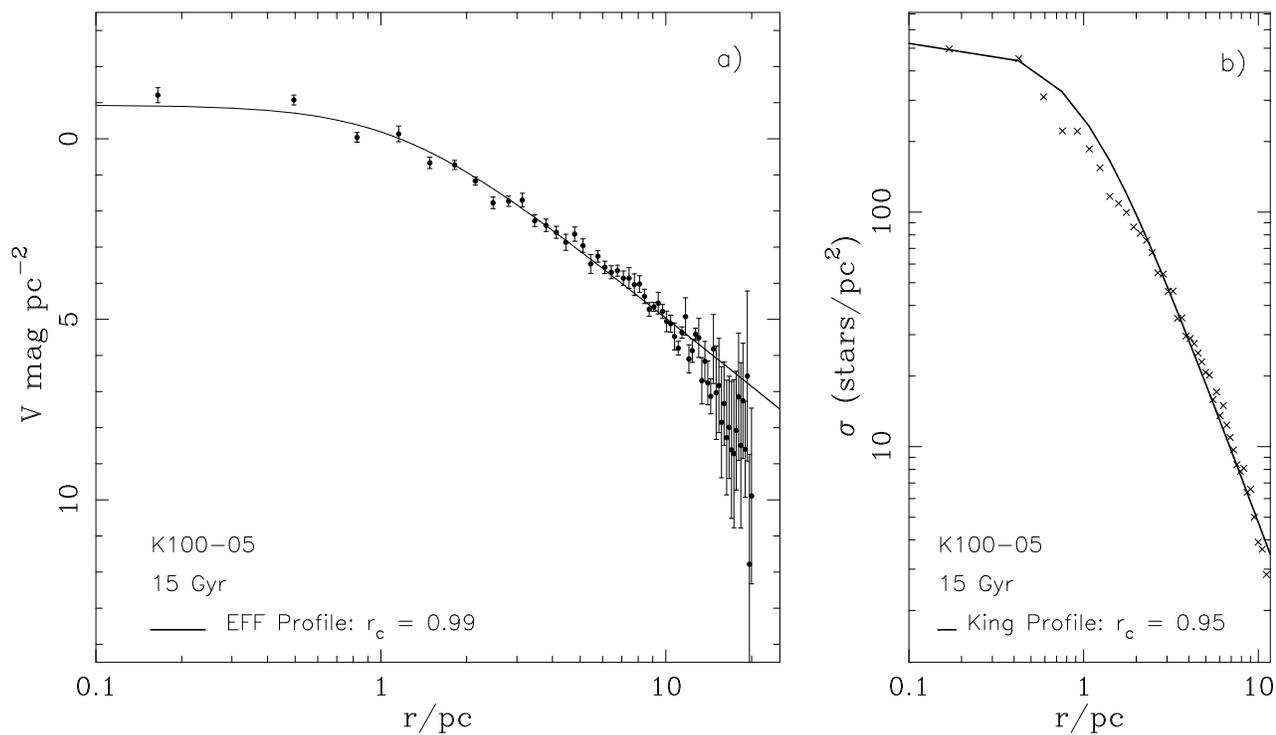}
\caption{
Demonstration of the fitting process used to determine the observational core radius, 
$r_{\rm c,l}$, using the K100-05 model at an age of $15\,$Gyr as an example. 
Shown are: a) the V magnitude surface brightness profile with the best fit 
Elson, Fall \& Freeman (1987) model; and, b) the surface density profile with the best fit 
King (1966) model. 
In both cases the data are projected along the Y-axis and stars fainter than $M_{\rm V} = 10$ 
or more than two magnitudes brighter than the main-sequence turn-off are excluded. 
\label{f:fig3}}
\end{figure*}

As an example the SBP and EFF model fit for the K100-05 simulation at $15\,$Gyr 
is shown in Figure~\ref{f:fig3}a. 
The resulting core radius is $r_{\rm c,l} = 0.99\,$pc. 
For comparison Figure~\ref{f:fig3}b shows the projected surface density profile of the 
same stars along with the best fitting King model (King 1966). 
This gives $r_{\rm c,l} = 0.95\,$pc in good agreement. 
The corresponding $N$-body core radius for the model cluster is $0.4\,$pc. 
Values of $r_{\rm c}$, $r_{\rm h}$, $r_{\rm c,l}$ (from EFF) and $r_{\rm h,l}$ at $15\,$Gyr 
for each simulation are given in Table~\ref{t:table1}. 

\begin{figure*}
\includegraphics[width=168mm]{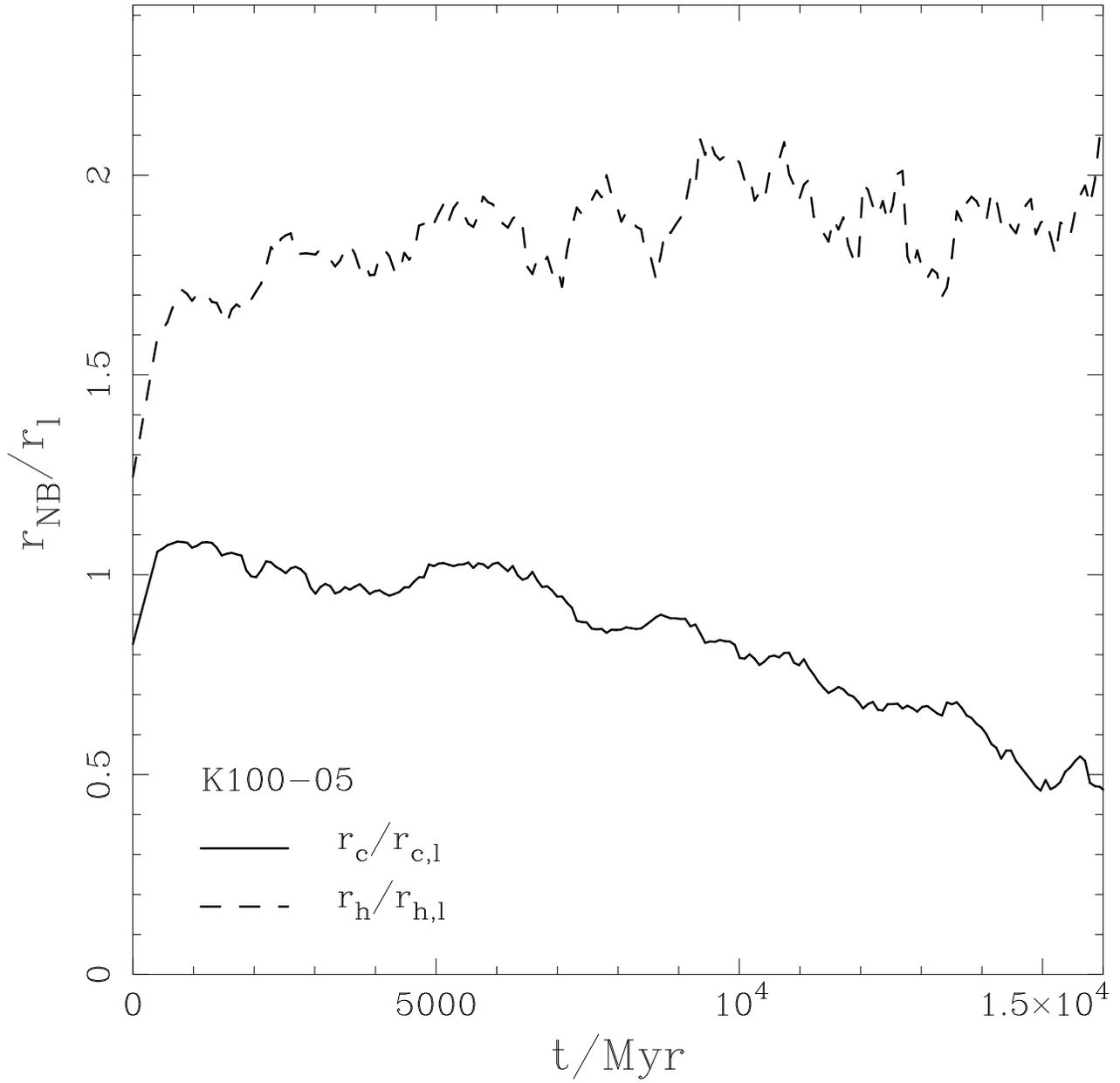}
\caption{
Comparison of radii calculated using the standard $N$-body method to those 
calculated from fitting to the simulated luminosity profiles. 
Data from the K100-05 simulation starting with $95\,000$ single stars and $5\,000$ binaries.
\label{f:fig4}}
\end{figure*}

Figure~\ref{f:fig4} demonstrates the relationship between the $N$-body and observationally 
determined radii for the K100-05 simulation as it evolves. 
For the most part $r_{\rm h} \simeq 2 \, r_{\rm h,l}$ in agreement with 
Baumgardt, Makino \& Hut (2005). 
It is important to emphasize that $r_{\rm h,l}$ is derived from a 2-dimensional projection of the 
$N$-body data whereas $r_{\rm h}$ is based on the original 3-dimensional data. 
Simply calculating $r_{\rm h}$ from a 2-dimensional projection gives a reduction of 
about 25\%, as expected (see Fleck et al. 2006), and 
using the stellar light gives a further reduction. 
It can also be seen from Figure~\ref{f:fig4} that for the first $\sim 7\,$Gyr $r_{\rm c}$ is a good 
approximation to $r_{\rm c,l}$. 
However, as the cluster becomes dynamically old ($ t > 5 \, t_{\rm rh,0}$) this 
approximation is no longer valid. 
During core-collapse the central density increases and the value of $r_{\rm c}$ 
computed from the density-weighted procedure decreases (as witnessed in Figure~\ref{f:fig2}). 
At the same time the remnant fraction in the core is increasing (Baumgardt \& Makino 2003) 
which flattens the profile of the visible stars and causes $r_{\rm c,l}$ to be greater than 
$r_{\rm c}$. 

After repeating the SBP-fitting process for the full set of simulations 
Figure~\ref{f:fig2} is repeated but now using $r_{\rm c,l}$ and $r_{\rm h,l}$. 
The result is shown in Figure~\ref{f:fig5}. 
This ratio, $\left( r_{\rm c} / r_{\rm h} \right)_{\rm l}$, can be compared 
to observational data. 
It is clearly evident that the ratio is higher than previously reported 
-- at $15\,$Gyr $r_{\rm c} / r_{\rm h} \sim 0.3$ regardless of binary 
content and without invoking an IMBH. 
Also plotted in Figure~\ref{f:fig5} is a fourth simulation, K100-00b. 
The setup for this model was identical to that of K100-00a except for the 
seed of the random number generator. 
However, unlike K100-00a this alternate model formed a BH-BH binary 
in the core after $4\,$Gyr of evolution. 
The BH masses are $24$ and $25 M_\odot$ and the binary formed in a 3-body 
interaction with an initial period of $19\,000\,$d. 
At $16\,$Gyr it was still present in the core with a period of $195\,$d. 
The energy generated in 3-body encounters between this binary and stars 
in the core acts to `puff-up' the core and inflate the core radius. 
This is analogous to what Baumgardt, Makino \& Hut (2005) find when 
an IMBH is present in the core. 
The K100-00b simulation maintains $r_{\rm c} / r_{\rm h} \sim 0.6 - 0.7$ throughout 
the evolution 
and is clearly distinct from the other models from about $11\,$Gyr onwards. 

For comparison, 
the first long-lived binary in the K100-00a simulation also formed at about $4\,$Gyr 
-- comprised of a white dwarf and a helium star -- but this simulation did not 
form a BH-BH binary at any point. 
The reason for this is related to the velocity kicks given to supernovae remnants. 
For the models in this work, when a neutron star or BH is born a velocity kick 
chosen at random from a uniform distribution between $0 - 100\, {\rm km} {\rm s}^{-1}$ 
is applied. 
This leads to retention fractions of 15-20\% which is in line with the suggestions 
of Pfahl, Rappaport \& Podsiadlowski (2002) for GCs. 
The initial K100-00a model contained 39 main-sequence stars with mass in excess of $20 M_\odot$, 
i.e. stars that would evolve to form BHs. 
However, only five BHs were retained in the model cluster after birth and only one 
of these BHs had a mass in excess of $20 M_\odot$. 
The K100-00b model started with 42 massive main-sequence stars and had eight retained BHs, 
three of which where more massive than $20 M_\odot$. 
So the velocity kick process, which itself is uncertain, 
and the related small number statistics of BH numbers, 
are certainly playing a role in determining the evolution histories of the models.

In Figure~\ref{f:fig6} the projected surface density profiles of the 
K100-00a, K100-00b and K100-05 simulations at $15\,$Gyr are compared. 
Profiles are constructed using radial bins of $500$ stars each and all stars are included. 
The profile of the K100-10 model is similar to that of the K100-05 model. 
Comparison of the K100-00a and K100-05 profiles shows the expected result 
in that the single star model is more centrally condensed and would return a smaller 
core radius from King model fitting. 
By contrast the K100-00b profile is much flatter. 
Thus the behaviour seen for $N$-body models with a central IMBH 
(Baumgardt, Makino \& Hut 2005) can be replicated by the presence of 
a central stellar mass BH-BH binary. 

Finally, the K100-00a and K100-10 simulations are used to look at the effect of 
primordial binaries on the distribution of remnants in evolved clusters. 
Baumgardt et al. (2003a) showed that the density profile of remnants 
(white dwarfs, neutron stars and stellar-mass BHs) rises more strongly in the centre 
of a cluster than the profile of luminous, or observable, stars. 
Thus the mass-to-light ratio rises naturally towards the centre of a cluster without 
the need for an IMBH. 
This point was shown by Baumgardt et al. (2003a) to be important when interpreting 
the observed velocity dispersion profile of M15 which had been used to infer the 
presence of an IMBH in the core (Gerssen et al. 2002). 
The Baumgardt et al. (2003a) models did not include primordial binaries. 
Thus the K100-00a model in this work can be expected to show similar behaviour. 
Figure~\ref{f:fig7}a shows the projected density profiles of remnant stars 
and luminous stars (main-sequence stars with $M_{\rm V} < 10$ and giants) at $15\,$Gyr. 
Indeed the remnant profile of model K100-00a rises more steeply towards the centre.  
For this model at $15\,$Gyr remnants comprised 40\% of the cluster mass but only 1\% of 
this was in the form of neutron stars and BHs. 
Note that to probe deeper into the centre of the model clusters 
$100$ stars per bin has been used in Figure~\ref{f:fig7} which explains why the profiles are more 
erratic than those of Figures~\ref{f:fig3}b and \ref{f:fig6}. 
In Figure~\ref{f:fig7}b this exercise is repeated for the K100-10 model 
-- the presence of 10\% primordial binaries has erased any difference between the profiles. 
This is because binaries present a population of comparable average mass to the 
remnants and therefore segregate towards the centre on a similar timescale.

\section{Discussion}
\label{s:discus}

This work has gone some way to fulfilling a need identified by Trenti (2006) 
-- taking realistic $N$-body models and analyzing the snapshot data as if it 
were data acquired by a telescope. 
While these models are comparable in size to GCs at the lower end of 
the GC mass function (e.g. Gnedin \& Ostriker 1997) they should not be taken as directly 
applicable to GCs. 
The results presented are mainly for comparison to other $N$-body models 
-- they provide an excellent companion to the models of 
Baumgardt \& Makino (2003) and Baumgardt, Makino \& Ebisuzaki (2004) 
and a step forward in particle number compared to Trenti, Heggie \& Hut (2007). 
Good agreement is also found with the Monte Carlo models of cluster evolution 
performed by Fregeau \& Rasio (2007). 
Across a series of models starting with $100\,000$ objects (stars and binaries) 
these authors report $r_{\rm c} / r_{\rm h}$ values in the range $0.05 - 0.1$ 
with little to no dependence on the initial cluster profile or binary fraction. 
This is the same range shown for the $N$-body models in Figure~\ref{f:fig2}, 
noting that the Monte Carlo $r_{\rm c} / r_{\rm h}$ is calculated using the 
traditional $N$-body method. 
Significantly, Fregeau \& Rasio (2007) do not see any noticeable change in 
$r_{\rm c} / r_{\rm h}$ when they move to models starting with $300\,000$ objects. 

Trenti (2006) looked at globular cluster data from the catalogue of 
Harris (1996\footnote{The updated version of this catalogue is available 
on-line at either http://physwww.mcmaster.ca/\%7Eharris/mwgc.dat 
or http://coihue.rutgers.edu/~andresj/gccat.html}) 
to examine the distribution of $r_{\rm c} / r_{\rm h}$ ratios. 
This involved carefully selecting a sample of 57 GCs with the main determinant 
being that the measured half-mass relaxation timescale be less than $10^9\,$yr. 
This was to ensure that the GCs in the sample were all dynamically old 
-- at least 10 half-mass relaxation times old based on a conservative age 
estimate of $10\,$Gyr. 
The motivation for doing this was based on the demonstration by Trenti (2006) 
that $r_{\rm c} / r_{\rm h}$ can only be used to distinguish between clusters 
with differing initial content (single stars, primordial binaries and IMBHs) after 
at least 10 half-mass relaxation times have elapsed. 
However, this model result was based on the use of the initial half-mass relaxation 
timescale ($t_{\rm rh,0}$) whereas the observed clusters only give the current $t_{\rm rh}$. 
As has been shown in this work,  
this can be expected to be at least a factor of two less than $t_{\rm rh,0}$. 
Another consideration is that $t_{\rm rh}$ calculated from models is using the 
three-dimensional half-mass radius while the value of $t_{\rm rh}$ quoted in the 
Harris (1996) catalogue is based on the two-dimensional half-light radius. 
This can also lead to an overestimate of the true dynamical age of a cluster\footnote{
This point came to mind after noting the conversion applied in 
Baumgardt, Makino \& Hut (2005) when calculating the relaxation timescale for 
Galactic GCs}. 
Thus caution is urged when comparing dynamical ages of model and real clusters. 

\begin{figure*}
\includegraphics[width=168mm]{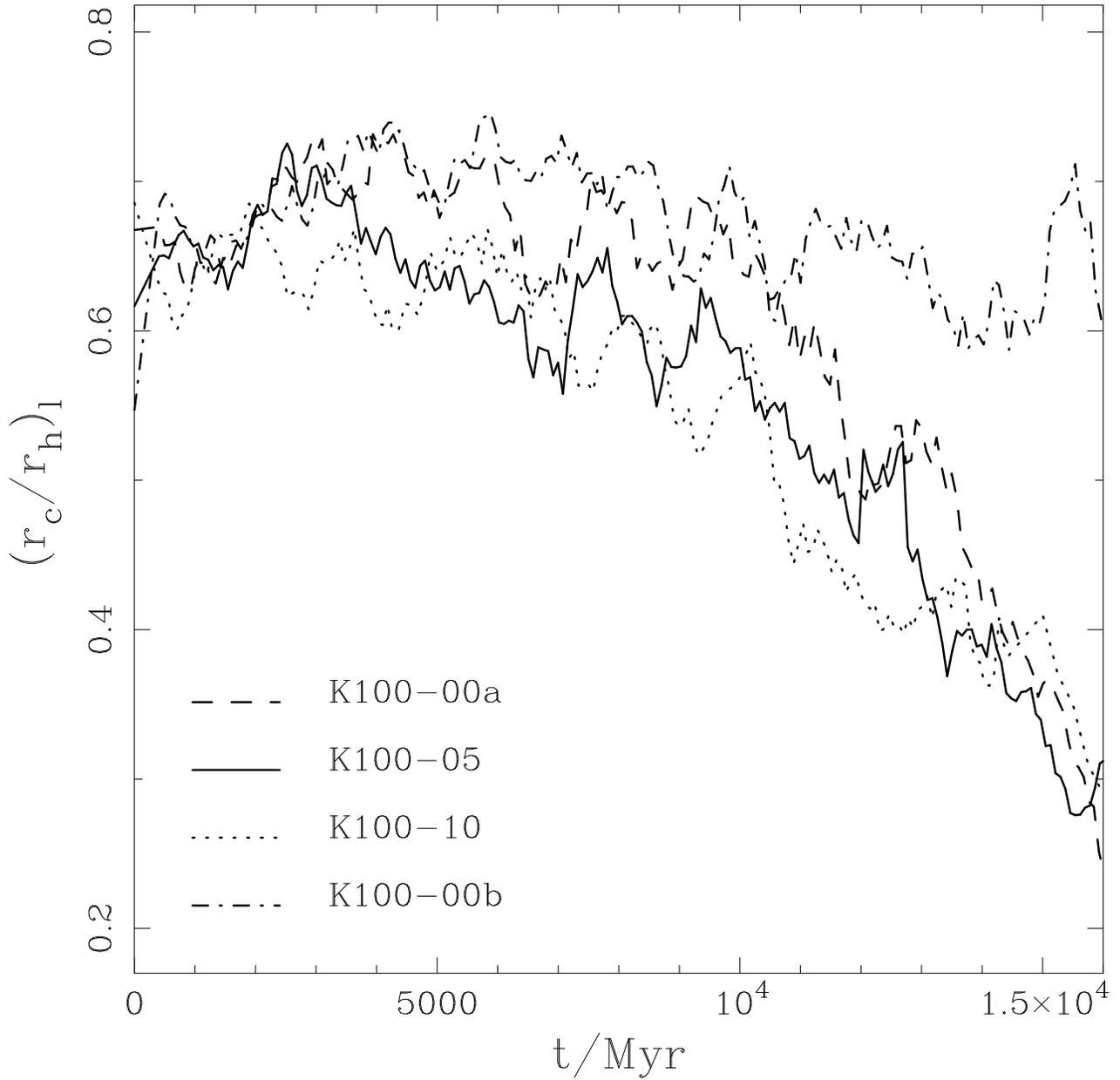}
\caption{
Evolution of the ratio of the core-radius, $r_{\rm c,l}$, to half-light radius, $r_{\rm h,l}$, 
for models starting with 0, 5 and 10\% binaries (see Table~\ref{t:table1} for a description). 
The radii are calculated from the simulated luminosity profiles using two-dimensional 
projected data (see text for details). 
An additional model that started with 0\% binaries but formed a BH-BH binary at $4\,$Gyr 
is included (K100-00b). 
\label{f:fig5}}
\end{figure*}

\begin{figure*}
\includegraphics[width=168mm]{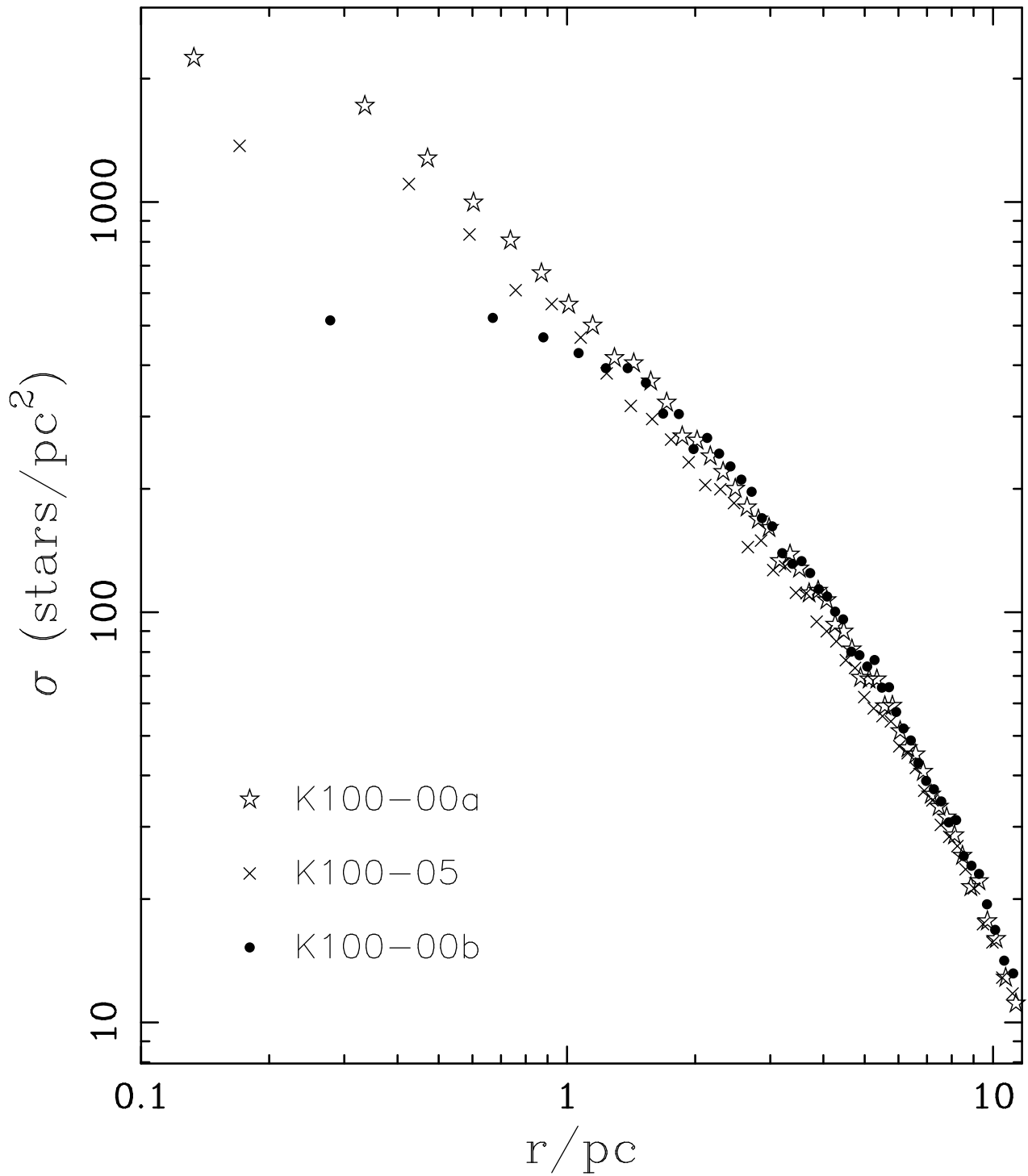}
\caption{
Projected surface density profiles at $15\,$Gyr for the K100-00a, K100-05 and K100-0b models. 
All stars are included and the profiles are computed using radial bins of 500 stars. 
\label{f:fig6}}
\end{figure*}

\begin{figure*}
\includegraphics[width=168mm]{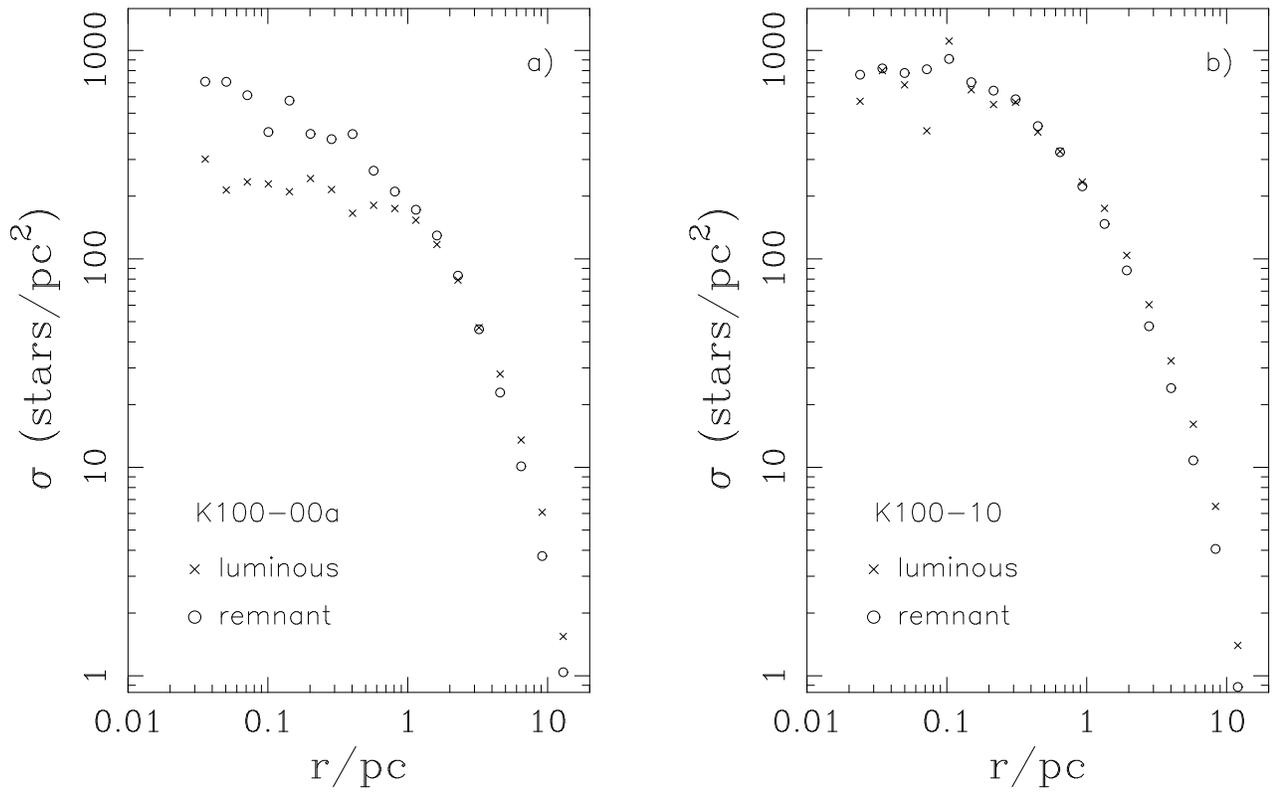}
\caption{
Comparison of the projected density profiles for remnant (circles) and bright (crosses) stars 
in: a) the K100-00a model at $15\,$Gyr; and, b) the K100-10 model at $15\,$Gyr. 
\label{f:fig7}}
\end{figure*}

In fact, Trenti (2006) also considered a refined sample based on 
a criterion of $t_{\rm rh} < 0.5 \times 10^9\,$yr which is more 
appropriate in terms of ensuring the clusters are dynamically old. 
This led to a sample of 25 Galactic GCs. 
Of these there are 10 with $r_{\rm c} / r_{\rm h} > 0.2$ which is the condition used 
by Trenti (2006) to infer the possible presence of an IMBH. 
The larger sample considered by Fregeau et al. (2003: see their Fig. 17) shows 
that in general a Galactic GC is as likely to have $r_{\rm c} / r_{\rm h} > 0.2$ than not. 
An important point here is that Fregeau et al. (2003) and Trenti (2006) are using 
$r_{\rm c} / r_{\rm h}$ from models (corresponding to Figure~\ref{f:fig2}) and 
this will underestimate the true ratio when compared to observations which use 
$r_{\rm c,l} / r_{\rm h,l}$ (as given in Figure~\ref{f:fig5}). 
The results presented here show that observed ratios up to at least 0.4 do not require an IMBH 
for explanation. 
The median $r_{\rm c} / r_{\rm h}$ for the Trenti (2006) sample is 0.28 which is well 
matched by the $N$-body models. 
As a result it is suggested, conservatively, that $r_{\rm c} / r_{\rm h} > 0.5$ be used to distinguish 
clusters which require the presence of something out of the ordinary to explain their 
inner structure. 
In the Harris (1996) database there are six such clusters (with $t_{\rm rh} < 0.5 \times 10^9\,$yr). 
They are E3, Terzan 3, NGC$\,6366$, Pal 6, NGC$\,6535$ and Pal8. 

The $N$-body models have shown that one explanation for a cluster observed to have a 
large $r_{\rm c} / r_{\rm h}$ 
ratio is the presence of a stellar mass BH-BH binary. 
The action of this BH-BH binary causes $r_{\rm c} / r_{\rm h}$ to diverge from that 
found in models without such a binary. 
This divergent behaviour occurs after about $11\,$Gyr, in terms of model age 
(see Figure~\ref{f:fig5}), which in dynamical terms equates to approximately six 
half-mass relaxation times (see Figure~\ref{f:fig1}). 
So it is possible to differentiate between models with and without a BH-BH binary 
before the completion of core-collapse evolution. 

The findings relating to the BH-BH binary model (K100-00b) occurred very much by chance 
as this study in no way set out to create a model that would form such a binary. 
What it does demonstrate is that the random meeting of stars in a cluster (model or real), 
and additionally the randomness introduced by velocity kicks given to supernova remnants, 
can have severe implications for the long-term structure and observed nature of a cluster. 
This particular model also showed a flattened density profile compared to the models 
that did not form a long-lived central BH-BH binary. 
Baumgardt, Makino \& Hut (2005) found that models with an IMBH also gave flattened 
profiles. 
Comparison to the observed surface brightness profiles of 37 Galactic GCs presented 
by Noyola \& Gebhardt (2006) lead to the suggestion of five clusters of interest in 
terms of detecting an IMBH. 
These clusters may also be of interest for finding a BH-BH binary. 

The results in this study have actually made it more difficult to explain clusters with low 
$r_{\rm c} / r_{\rm h}$ 
-- approximately half of the Galactic GCs have ratios less than 0.2 
(Harris 1996; Fregeau et al. 2003; Trenti 2006) and 
these cannot be reached by the models (based on inspection of Figure~\ref{f:fig5}). 
However, there are two factors to note here. 
The first is that the results shown in Figure~\ref{f:fig5} are smoothed. 
Looking at the raw, non-smoothed, data there is much fluctuation and values 
below 0.2 do occur in the models at late times (excluding the K100-00b model) 
-- the average error in the smoothed $\left( r_{\rm c} / r_{\rm h} \right)_{\rm l}$ 
as shown in Figure~\ref{f:fig5} is approximately $\pm 10\%$. 
So small values are certainly possible depending on when a cluster is `observed'. 
A word of caution is required on this point as real GCs are, in most cases, richer than 
the models presented here and statistical fluctuations will be smaller. 
The second point is uncertainty in the $r_{\rm c,l}$ fitting process. 
To demonstrate this one can look at the surface density profile of model K100-05 
at $15\,$Gyr, as shown in Figure~\ref{f:fig3}a. 
The King model fit shown gives $r_{\rm c} = 0.95\,$pc however, if the fitting process 
is biased to fit the inner $1\,$pc of the profile then values as low as $r_{\rm c} = 0.7\,$pc  
are plausible. 
It is also true that many of the GCs with $r_{\rm c} / r_{\rm h} < 0.2$ in the Harris (1996) 
catalogue are also flagged as core-collapse clusters and accurate measurement of 
$r_{\rm c,l}$ using SBP-fitting software can be difficult for clusters passing through this 
phase. 
The interested reader can look at the SBP library in Trager, King \& Djorgovski (1995) 
along with the associated description of the fitting process for core-collapse clusters. 

One could also expect that reducing the strength of the tidal field would lead 
to a reduction in $r_{\rm c} / r_{\rm h}$ through an increase in $r_{\rm h}$. 
However, there is no evidence in the Galactic GC sample for a link 
between $r_{\rm c} / r_{\rm h}$ and distance from the Galactic centre. 
Note also that the models of Baumgardt, Makino \& Hut (2005) that 
are of comparable size to those presented here, but with an IMBH in the core, 
actually give smaller $r_{\rm c} / r_{\rm h}$ values. 
One difference between the two sets of models is that the Baumgardt, Makino \& Hut (2005) 
models are isolated and indeed they do show a larger half-light radius. 
Even after correcting for this, the Baumgardt, Makino \& Hut (2005) models with an IMBH 
would give comparable (not larger) $r_{\rm c} / r_{\rm h}$ values to the models in this work 
without an IMBH. 
So it is not clear from this comparison that clusters with an IMBH should necessarily 
show a larger $r_{\rm c} / r_{\rm h}$ ratio, as suggested by the models described in 
Trenti (2006) and theoretical arguments (Heggie et al. 2006). 
It is interesting to note an apparent discrepancy between the $r_{\rm c} / r_{\rm h}$ values 
reported from the IMBH models of Trenti et al. (2007) and those of 
Baumgardt, Makino \& Hut (2005). 
Modelling time-dependent tidal fields may also be important in determining the 
actual $r_{\rm c} / r_{\rm h}$ ratio, and the choice of initial conditions, such 
as the scale radius, may also play a role.  
Clearly there is more work to be done in this field before we can resolve the issue 
of which GCs may harbour an IMBH. 

On one hand this investigation is suggesting that the presence of IMBHs in GC 
cores is not so likely 
-- intermediate $r_{\rm c} / r_{\rm h}$ values can be explained by models 
without an IMBH provided the correct comparison is made and higher 
$r_{\rm c} / r_{\rm h}$ values may instead show the presence of a stellar-mass 
BH-BH binary. 
However, the models also provide an opportunity to look at how the distribution 
of stars in an old cluster is affected by the presence of a sizeable primordial 
binary population. 
The model with 10\% primordial binaries shows that the mass distribution follows 
the light distribution throughout the cluster 
-- the steeper density profile of remnant stars compared to bright stars 
seen in the centre of single-star models is not replicated. 
Therefore, observations that infer a steepening mass-to-light ratio in the core of a 
globular cluster  should not be 
dismissed as a possible IMBH indicator (see also Gebhardt, Rich \& Ho 2005). 
This is provided we assume that globular clusters are born with a modest binary fraction 
(Hut et al. 1992). 
Exactly how a direct model of a GC with primordial binaries and an IMBH 
will behave is beyond the scope of this work.

\section{Summary}
\label{s:summ}

By treating model data as if it were observational data higher 
$r_{\rm c} / r_{\rm h}$ values than previously reported have been revealed. 
This provides a good match to the majority of the Galactic GCs without the need 
for an IMBH. 
It has also been shown that factors such as the presence of a BH-BH binary 
(comprised of stellar mass BHs) in a cluster core can flatten the measured 
luminosity profile and inflate the measured core-radius. 
None of this precludes the existence of IMBHs in GC cores. 
However, it does demonstrate that the $r_{\rm c} / r_{\rm h}$ ratio cannot 
be used with any certainty to infer the dynamical history or content of 
a cluster core.

\section*{Acknowledgments}

It is a pleasure to thank Dougal Mackey for providing the surface brightness 
profile software and Douglas Heggie for many helpful suggestions that 
improved the text. 
JRH thanks the Swinburne Research Development Scheme for travel support 
and the American Museum of Natural History for hosting a visit during this work.

\clearpage

\clearpage

\clearpage

\clearpage

\clearpage

\clearpage

\clearpage

\clearpage

\begin{table*}
\begin{minipage}{126mm}
\caption{
Parameters of the simulations performed in this work. 
Column~1 gives the label assigned to each model while Columns~2 and 3 
show the starting number of single stars and binaries, respectively. 
Columns~4-7 give the following radii (in pc) for the models at an age of $15\,$Gyr: 
the $N$-body density-weighted core radius; 
the half-mass radius; 
the EFF-fitted core radius; and, 
the half-light radius. 
Note that $r_{\rm c,l}$ and $r_{\rm h,l}$ are from two-dimensional projected data 
while $r_{\rm c}$ and $r_{\rm h}$ are based on three-dimensional data. 
\label{t:table1}
}
\begin{tabular}{lrrrrrr}
\hline 
Label & $N_{\rm s}$ & $N_{\rm b}$ & $r_{\rm c}$ & $r_{\rm h}$ & $r_{\rm c,l}$ & $r_{\rm h,l}$ \\
\hline
K100-00a & $100\,000$ &           $0$ & 0.34 & 4.89 & 0.85 & 2.34 \\
K100-00b & $100\,000$ &           $0$ & 1.27 & 5.59 & 1.88 & 3.72 \\
K100-05   & $  95\,000$ &   $5\,000$ & 0.40 & 5.25 & 0.99 & 2.75 \\
K100-10   & $  90\,000$ & $10\,000$ & 0.48 & 5.31 & 0.86 & 2.71 \\
\hline
\end{tabular}
\end{minipage}
\end{table*}

\label{lastpage}

\end{document}